\documentclass[smallextended]{svjour3}
\usepackage{amsmath}
\usepackage{amssymb}
\usepackage[T1]{fontenc}
\usepackage[latin1]{inputenc}
\usepackage{graphicx}
\usepackage{wasysym}
\usepackage{natbib}
\usepackage{lineno}
\usepackage[usenames]{color}
\usepackage{pdflscape}
\usepackage{rotating}
\usepackage{soul}
\usepackage{xcolor, colortbl}
\definecolor{Gray}{gray}{0.9}
\usepackage{etoolbox}
\usepackage[letterpaper,margin=2.5cm]{geometry}

\usepackage{tabularx}

\usepackage[pdftex,colorlinks,linkcolor=blue,anchorcolor=blue,urlcolor=blue,citecolor=blue]{hyperref}

\newtoggle{supplement}\toggletrue{supplement} 
\newtoggle{removemanunobox} \toggletrue{removemanunobox}

\newcommand{\E}{\mathbb{E}}

\iftoggle{removemanunobox}
{
\makeatletter
\renewcommand\makeheadbox{{%
\hbox to0pt{\vbox{\baselineskip=10dd\hbox
to\hsize{\kern3pt\vbox{\kern12pt
\hbox{\ }
\hbox{}
}}}%
}}}
\makeatother
}

\journalname{Climatic Change}

\title{Allowances for evolving coastal flood risk under uncertain local sea-level rise}
\author{Maya K. Buchanan, Robert E. Kopp, Michael Oppenheimer, Claudia Tebaldi}

\begin{document}

\institute{
M. K. Buchanan \at Woodrow Wilson School of Public \& International Affairs, Princeton University, Princeton, NJ, USA. \email{mayak@princeton.edu}
\and R. E. Kopp \at Department of Earth \& Planetary Sciences, Rutgers Energy Institute, and Institute of Earth, Ocean, \& Atmospheric Sciences, Rutgers University, New Brunswick, NJ, USA. 
\and M. Oppenheimer \at Woodrow Wilson School of Public \& International Affairs, \& Department of Geosciences, Princeton University, Princeton, NJ, USA.
 \and  Claudia Tebaldi \at Climate and Global Dynamics, National Center for Atmospheric Research (NCAR), Boulder, CO, USA}

\date{}
\maketitle \vspace{-12pt}

\begin{abstract}
Estimates of future flood hazards made under the assumption of stationary mean sea level are biased low due to sea-level rise (SLR). However, adjustments to flood return levels made assuming fixed increases of sea level are also inadequate when applied to sea level that is rising over time at an uncertain rate. SLR allowances---the height adjustment from historic flood levels that maintain under uncertainty the annual expected probability of flooding---are typically estimated independent of individual decision-makers' preferences, such as time horizon, risk tolerance, and confidence in SLR projections. We provide a framework of SLR allowances that employ complete probability distributions of local SLR and a range of user-defined flood risk management preferences. Given non-stationary and uncertain sea-level rise, these metrics provide estimates of flood protection heights and offsets for different planning horizons in coastal areas. We illustrate the calculation of various allowance types for a set of long-duration tide gauges along U.S. coastlines. 
%Our work is particularly relevant for the U.S. where federal flood protection standards are built upon the convention of the historic 100-year flood. However, we believe our methodology of providing actionable science constructed for heterogenous decision-making contexts is broadly applicable.

%To accommodate both the temporal dynamics of SLR and their uncertainty, we develop an Average Annual Design Life Level (AADLL) metric and associated Design Life SLR (DL-SLR) allowances. The AADLL is the flood level corresponding to a time-integrated annual expected probability of occurrence (AEP) under uncertainty over the design life of an asset; DL-SLR allowances are the adjustment from 2000 levels that maintain current average probability over the design life. 
\end{abstract}

\section{Introduction}

The distribution of coastal flood events is influenced by astronomical tides, the distribution of storm events, and local mean sea level \citep{Lin2012a, Hunter2012a}. Under current practice, acceptable levels of coastal flood risk are often based upon specific flood return periods, such as the 100-year flood (1\% annual expected probability of occurrence [AEP]) for the U.S. National Flood Insurance Program \citep{Galloway2006a}. While federally designated flood zones and often capital projects are based on flood probabilities that assume stationary sea level, sea-level rise (SLR) renders estimates of flood hazard exceedingly optimistic. 

For example, \citet{Talke2014a} show that stationary predictions of flood return levels fail to capture the rapidly increasing flood recurrence due to sea-level rise in Manhattan. The New York City Special Initiative for Rebuilding and Resiliency \citep{SIRR2013a} assessed how frequently elevated flood return levels would top the NYC subway system protection level with sea-level rise (using a 90th percentile SLR estimate of 31 inches by 2050). They found that this threshold---not surpassed until Hurricane Sandy in 2012---would be susceptible to a 25\% AEP flood. Similarly, since $\sim$50 cm of sea-level rise would increase the AEP of the current 0.1\% annual chance flood to 1\% at London's Thames Barrier \citep{Conner2013a}, the barrier---originally built in the 1970s to protect against the 1\% AEP flood---now faces a premature upgrade \citep{EnvironmentAgency2012a}. \citet{Houser2015a} estimate that, in the absence of adaptation, changes in flood frequency driven by SLR would cost about 20--30 billion dollars per year in the U.S. by the end of the century (assuming current economic valuation). Aware of these growing risks, U.S. cities and states are calling for metrics to help identify how much to adapt to this threat of uncertain magnitude \cite[e.g.,][]{Bierbaum2014a, Boston2014a}.

%It seems like there are two points being made in this paragraph that should be separated: One is that sea-level rise is uncertain; the other is that it is deeply uncertain. It's also confusing to introduce Hunter's allowances, and then jump back to deterministic offsets. It seems like you should present the 
%deterministic scenarios first, then present allowances as a way of dealing with uncertainty, and then introduce deep uncertainty.

%I donÕt think you are clear enough about what Hunter did and how we are different.  Hunter's work is partially probabilistic (in the IPCC sense, not full distributions) and, if you include his 2013 paper (not cited), does local SLR + surge.  The way it's written at beginning and end, I donÕt think this is clear.

%deterministic
While some authors have developed estimates of the changes in flood levels under the influence of SLR, adjustments to stationary AEPs made assuming fixed sea-level increases \citep[e.g.,][]{Tebaldi2012a} are inadequate when applied to sea level that is rising at an uncertain rate over time. Some studies have assumed fixed sea-level increases derived from deterministic scenarios of SLR that are not conditional upon emissions scenarios \citep{Parris2012a, USGCRP2014a, Kunkel2015a}. For example, in the Hurricane Sandy Tool Kit---a prominent sea-level rise adaptation tool for some Sandy-affected areas---New Jersey users are directed to follow the `high' federally vetted SLR scenario (2.0 m of global mean SLR by 2100) if their asset has low risk tolerance and the `low' SLR scenario (0.2 m rise by 2100) for high risk tolerance \citep{USGCRP2014a}. Such deterministic scenarios may be insufficient to capture the uncertainty of local SLR and its implications for local flood risk management. Additionally, the majority of studies have employed global mean sea levels, while others have accounted for some but not all local factors to provide local flood estimates (see supplemental information [SI] for a comparison of methods).

To help account for the uncertainty in SLR projections, \citet{Hunter2012a} developed the concept of SLR allowances---the vertical buffer necessary to maintain an AEP---estimated by global mean SLR (and later local SLR;  \citealt{Hunter2013a}) plus a margin for uncertainty provided by various parametric probability distribution functions (PDFs) by a fixed date (2100). This method provides an amount of freeboard for decision-makers to maintain their flood risk tolerance, using the Gumbel extreme value distribution to fit annual flood exceedances. Hunter's (\citeyear{Hunter2012a,Hunter2013a}) SLR allowances are for single time points (hereafter, `instantaneous allowances').

Although it is certain that SLR is occurring and will continue \citep{Church2013a}, its rate remains deeply uncertain and ambiguous \citep{Kasperson2008a, Heal2014a, Ellsberg1961a}, in the sense that no single probability distribution function (PDF) is widely accepted. This deep uncertainty poses a methodological challenge for integrating SLR projections into flood hazard characterization, and ultimately risk management. Moreover, individuals, businesses, and municipalities do not currently have systematic guidance regarding how much freeboard to account for SLR that reflects their managerial preferences. 

To help accommodate communities' need for resilience metrics, we combine four useful methods to expand upon Hunter's SLR allowances. First, we employ the temporally dynamic, uncertain SLR projections of \citet{Kopp2014a}, which provide reasonable, complete PDFs of local sea-level changes across a range of sites. Second, we address deep uncertainty by using a limited degree of confidence metric \citep{Froyn2005a, Mcinerney2012a}. %that reflects the limitations of deeply uncertain projections. 
Third, we provide an additional allowance type inspired by the work of Rootz\'en and Katz (\citeyear{Rootzen2013a}) for hydrologic design-life levels to accommodate different assets' lifetimes in non-stationary risk management. We define the average annual design-life level (AADLL) as the flood level corresponding to a time-integrated AEP under uncertainty over the lifetime of an asset, and we define the associated design-life (DL) allowance as the adjustments from historical levels that maintain historical probability of flooding over a given design life. Fourth, because building resilience into infrastructure also requires user- and asset-specific risk preferences \citep{Adger2009a}---such as flood risk tolerance, SLR risk perception, and valuation of asset protection---we incorporate these features into the design-life and instantaneous allowances.

In Section 2, we lay out the formal framework underlying AADLL and DL allowances and describe the calculation of historical flood return periods and of sea-level rise projections, and the treatment of uncertainty. Section 3 illustrates calculation of AADLL flood levels and DL allowances with a representative set of 71 long-recording tide gauges along U.S. coastlines. Section 4 discusses how these metrics might be applied in the context of sea-level rise resilience decision-making. Conclusions are presented in Section 5.

\section{Methods}
\subsection{Framework}
While \citet{Rootzen2013a} discussed design-life levels in the context of hydrological flood hazard analysis, the concept is equally applicable to other extremes, including extreme coastal flood heights. As defined by \citet{Rootzen2013a}, the $t_1 - t_2$ $p\%$ design-life level is the level of an extreme that has a $p\%$ probability of occurrence over the time period $t_1$ to $t_2$. We extend this concept by defining the Average Annual Design-Life Level (AADLL), which is more directly comparable to the AEPs and associated flood heights used in flood risk management. The $t_1 - t_2$ p\% AADLL has an average $p\%$ per year probability of occurrence over the interval $t_1 - t_2$. For example, in the context of coastal flood risk, the 2020--2050 $1\%$ AADLL is the flood height that has an average $1\%$ per year probability of occurrence over the 30 years between 2020 and 2050. Under stationary sea levels, the 1\% AADLL is equal to the height of the historic 1\% AEP flood. 

Expressed more formally, let $N(z)$ be the number of expected floods per year exceeding height $z$ under stationary sea level, which can be estimated from the application of extreme value theory to tide-gauge statistics. By definition, $1/N(z)$ is the return period of a flood of height $z$. For an arbitrary sea-level change $\Delta$, assuming no change in the distribution of flood heights relative to mean sea level, the number of expected floods of height $z$ is $N(z - \Delta)$. Letting the uncertain sea-level rise at time $t$ be denoted by $\Delta_t$, we define the instantaneous number of expected floods per year of height $z$ as
\begin{align}
N_{e}(z,t) & =  \E[N(z - \Delta_t)].
\end{align}
The average annual expected number of floods over period $t_1$ to $t_2$ is then given by 
\begin{align}
\tilde{N}_{e}(z,t_1,t_2) & =  \frac{1}{t_2 - t_1} \int_{t_1}^{t_2} N_e(z,t) dt. 
\end{align}
Accordingly, the $t_1 - t_2$ $p\%$ AADLL is the value of $z$ such that  $\tilde{N}_{e}(z,t_1,t_2) = p\%$. 

The instantaneous allowance of \cite{Hunter2012a} is defined as the level $A(N_0,t)$ such that
\begin{align}
N_{e}(z + A(N_0,t),t) & = N(z),
\end{align}
where $N_0 = N(z)$ is the number of expected floods in the absence of SLR. For example, $A(0.01,t)$ is the additional height above the current 1\% probability flood level needed to maintain an expected 1\% probability flood level at time $t$. 
(Note that, for known sea-level rise $\Delta$, $A = \Delta$.) Similarly, the DL allowance $\tilde A(N_0, t_1,t_2)$ is defined by
\begin{equation}
\tilde N_{e}(z + \tilde{A}(N_0,t_1,t_2),t_1,t_2) = N(z).
\end{equation}

\noindent For the Gumbel distribution of $N(z)$ assumed by Hunter, $A$ is independent of $N_0$, but this is not generally the case.

%If $N(z)$ follows a Generalized Extreme Value (GEV) Gumbel distribution, $A(N_0,t)$ (and therefore $\tilde A(N_0,t_1,t_2)$) is independent of $N_0$, as demonstrated by \citet{Hunter2012a} %(see SI); 
%this is not, however, generally true for a Poisson-Generalized Pareto Distribution (GPD), for which log Nz is not necessarily linear in Z. the curvature of $log N(z)$ differentiates between annual chance flood events (such as the 1\% and 0.2\% AEP) (Figure 1). 

%\begin{figure}[h!]
%\begin{center}
%\includegraphics[width=4.5in]{GPDvsGumbel.pdf} 
%\caption{Example illustration of the relationship between $N(z)$ and $z$ using the GPD vs Gumbel distribution.}
%\end{center}
%\end{figure}

%Whereas the Gumbel captures an exponential relationship between the annual expected number ($N(z)$) and height ($z$) of flood events (meters of storm tide above Mean High High Water (MHHW)). Because all terms are in the exponential, the change in the number of expected events is independent of annual chance flood levels. 

Calculating an AADLL thus requires both an estimate of the historic extreme value distribution $N(z)$ and a probability distribution of sea level over time, $P(\Delta, t)$. 

\subsection{Flood return levels under stationary sea level}
We use extreme value analysis (EVA) to assess flood return levels. EVA has commonly been used in engineering statistics since the 1950s to estimate the occurrence of extreme events, which by definition are too rare to be estimated by observations alone \citep{Coles2001a}. Using the GPD and peak-over-threshold (POT) approach to estimate local extreme water level exceedances, we estimate $N(z)$ for each tide gauge, following the methodology of \citet{Tebaldi2012a}. We analyze National Oceanic and Atmospheric Administration (NOAA) hourly tide-gauge records for sites with a minimum 30-year record (which can be found at http://tidesandcurrents.noaa.gov/; see SI for a list of record lengths). We consider 30 years to be the minimum required length for the trend not to exhibit significant multi-decadal cyclicity \citep{Tebaldi2012a}. A declustering routine isolates events that are spaced from each other by at least one day.
%To ensure that extremes are independently distributed, the tide-gauge data are declustered when multiple successive days are above the threshold, such that only one exceedance event, equal to the highest of the days, is counted.
%such that daily values are considered above the threshold only if they are separated by at least one day. 

Each record is linearly detrended to remove the effect of long-term sea-level rise and capture a distribution of exceedances influenced by sub-decadal sea-level variability, astronomical tides and storm surge alone. %The detrending is done to screen out the forced component of SLR at the gauge. 
We employed a linear trend rather than removing annual mean sea level because we wished to retain interannual sea-level variability in the extreme distribution.

The GPD takes the functional form
\begin{equation}
P(z - \mu \leq y|z > \mu) = \begin{cases}
1 - (1+ \frac{\xi y}{\sigma})^{\frac{-1}{\xi}} & \text{for } \xi \ne 0 \\
1 - \exp(-\frac{y}{\sigma}) & \text{for } \xi = 0
\end{cases}
\end{equation}
where $\mu$ is the water-level threshold above which exceedances are estimated, and $\sigma$ and $\xi$ are respectively the scale and shape parameters. The shape parameter $\xi$ controls the overall shape of the distribution's tail, with $\xi = 0$ giving rise to a Gumbel distribution, $\xi > 0$ giving rise to a heavier tailed distribution, and $\xi < 0$ to a bounded distribution. Assuming the probability of $z>\mu$ is Poisson-distributed with mean $\lambda$, the expected number of annual exceedances of height $z$ is given (for $z > \mu$) by
\begin{equation}
N(z) = \begin{cases}
\lambda (1+ \frac{\xi ( z - \mu)}{\sigma})^{\frac{-1}{\xi}} & \text{for } \xi \ne 0 \\
\lambda \exp(-\frac{ z - \mu}{\sigma}) & \text{for } \xi = 0
\end{cases}
\end{equation}
%We employ maximum-likelihood parameter estimates for $\mu$, $\sigma$ and $\xi$.

Compared to the GEV block maxima approach, the GPD POT approach has the advantage of extracting more information by using all of the data over the threshold (rather than just the yearly maxima), which improves the accuracy of the parameter estimates of the resulting distribution \citep{Coles2001a}. POT thresholds are set to accurately approximate the Poisson distribution of actual extreme outliers---high enough to justify the limiting distributional assumption of a GPD for the threshold exceedances, yet low enough to extract enough sample points to provide a reliable estimate of the parameters of the GPD. The diagnostics for the choice of the thresholds rely on the assessment of the behavior of the exceedances according to well-established metrics for the fitting of the parameters of GPDs \citep{Coles2001a, Tebaldi2012a}. A threshold equal to the 99th percentile of the distribution of daily maximum water levels (computed from hourly records) gave reasonable results for all of the tide gauges tested by \citet{Tebaldi2012a}. 

To account for parameter uncertainty, we estimate the maximum-likelihood shape and scale parameters and their covariance. Assuming the parameter uncertainty is normally distributed, we sample 1000 parameter pairs with Latin hypercube sampling. We then calculate the expected number of exceedances under parameter uncertainty, which we use for our main calculations. Sites' maximum-likelihood shape parameter values and historic 1\% AEP and 10\% AEP flood levels are shown in Figure 1. (See SI for parameter distributions for all sites). To allow our analysis to extend approximately to events with greater frequency, we assume that flood waters exceed mean higher-high water (MHHW) 182.6 times per year (i.e., every other day), and that events with frequency between $\lambda$ and 182.6/year are Gumbel distributed. We do not consider flood events more frequent than 182.6/year.

\subsection{Sea-level rise projections}
To estimate the time-varying probability of sea-level rise, we employ the local sea-level rise PDFs of \citet{Kopp2014a} for Representative Concentration Pathway (RCP) 8.5, which is frequently taken as a `business-as-usual' emissions pathway. %For each RCP, Kopp et al. produced a 10,000 member Monte Carlo ensemble of time series, which we employ directly in our analysis. 
\citet{Kopp2014a} constructed global sea level PDFs by combining global climate model (GCM) projections of thermal expansion, glacier surface mass balance model projections, semi-empirical projections of land water storage changes, and ice sheet projections based upon a combination of the expert assessment of the Intergovernmental Panel on Climate Change's Fifth Assessment Report and the expert elicitation study of \citet{Bamber2013a}. These global projections were localized by accounting for static-equilibrium fingerprint effects of land ice mass changes, GCM projections of atmosphere/ocean dynamics, and tide-gauge based estimates of non-climatic contributors to sea-level change, such as glacial-isostatic adjustment. From the probability distributions associated with each of these contributing factors, \citet{Kopp2014a} generated 10,000 samples of relative sea-level change at each of 1091 tide gauges. Kopp et al's (\citeyear{Kopp2014a}) median projected SLR from 2000 to 2100 under RCP 8.5 is illustrated in Figure 1b. We combined 10,000 Monte Carlo samples from the Kopp et al distributions of relative sea-level change with the extreme water level probability distributions to compute changes in flood return periods in response to SLR.

\begin{figure}[h!]
\begin{center}
\includegraphics[width=6.5in]{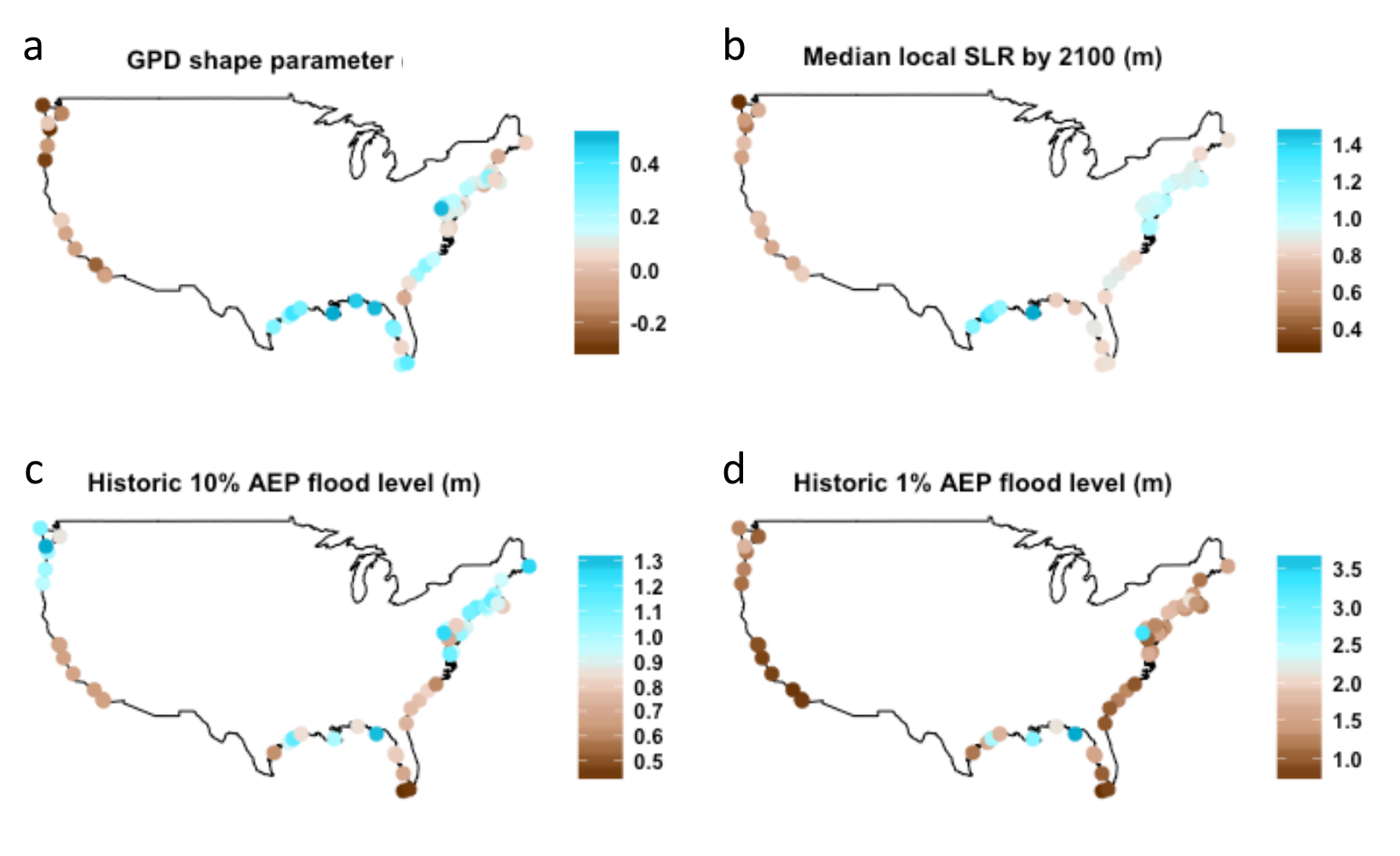} 
\caption{(a) Maximum-likelihood estimate of the GPD shape parameter, (b) median projected sea-level rise between 2000 and 2100 under RCP 8.5, (c) expected historic 10\% AEP flood level (meters above MHHW), and (d) expected historic 1\% AEP flood level for representative tide gauges (meters above MHHW).}
\end{center}
\end{figure}

\subsection{Ambiguity in sea-level rise projections}
The \citet{Kopp2014a} projections provide one plausible, self-consistent set of local sea-level rise PDFs, but they are not the only plausible PDFs. To accommodate imperfect confidence in these PDFs, we adapt the Limited Degree of Confidence (LDC) criterion used in decision-making under uncertainty \citep{Froyn2005a, Mcinerney2012a}. Taking $P(\Delta,t)$ from \citet{Kopp2014a}, we define the LDC effective probability as
\begin{equation}
\tilde{P}(\Delta,t) = \beta P(\Delta,t) + (1-\beta) \delta(\Delta - \Delta_{t,\mathrm{WC}})
\end{equation}
Here, $\beta \in [0,1]$ is a measure of confidence in $P(\Delta,t)$, $\Delta_{t,\mathrm{WC}}$ is a worst-case projection at time $t$, and $\delta$ is the Dirac delta function. %($\delta(0) = 1$, $\delta = 0$ otherwise). 
For $\Delta_{t,\mathrm{WC}}$, we adopt the 99.9th percentile projections of \cite{Kopp2014a}, which are comparable to other estimates of physically-plausible worst-case projections available in the literature \citep[e.g.,][]{Miller2013a, Pfeffer2008a, Sriver2012a}. It follows that
\begin{align}
N_{e,LDC}(z,t,\beta) & = \beta N_{e}(\Delta, t) + (1-\beta) N(z - \Delta_{t,\mathrm{WC}}) \\
\tilde{N}_{e,LDC}(z,t_1,t_2,\beta) &  =  \frac{1}{t_2 - t_1} \int_{t_1}^{t_2} N_{e,LDC}(z,t) 
\end{align}
Because of the extra weight given to the worst-case outcome, SLR allowances will be higher for decision-makers with incomplete confidence in the expert PDF than with full confidence.

\subsection{Combination of methods}
The allowance framework permits decision-makers to choose among several options based on their project and preferences. Figure 2a illustrates a simple flow chart of the combined framework's application. First, a decision-maker assesses her asset-specific flood protection and SLR preparedness preferences. Second, she selects the design life of her asset. Third, she selects an allowance type (DL or instantaneous). A DL allowance keeps annual risk below target in early years and above target in late years, while an instantaneous allowance for the end of the asset is more conservative, keeping annual risk below target throughout. Fourth, she selects a $\beta$ value to reflect her level of confidence in the expert PDF. Finally, she may wish to add a margin of safety to help protect against a potential increase in the number of coastal storms, which is a source of deeper uncertainty \citep{Christensen2013a, Church2013a}. For example, a homeowner in Boston may wish to elevate her structure so as to maintain her current 1\% AEP flood hazard over the lifetime of her mortgage, from 2020 to 2050 (Figure 2b). If she prefers to minimize her home's elevation to maintain her risk target on average over the period, as opposed to keeping her risk below target in all years except 2050, she selects the DL allowance. Finally, if she is fully confident in the \citet{Kopp2014a} local SLR PDF, her SLR allowance is 0.3 m. 

\begin{figure}[h!]
\begin{center}
\includegraphics[width=7.25in]{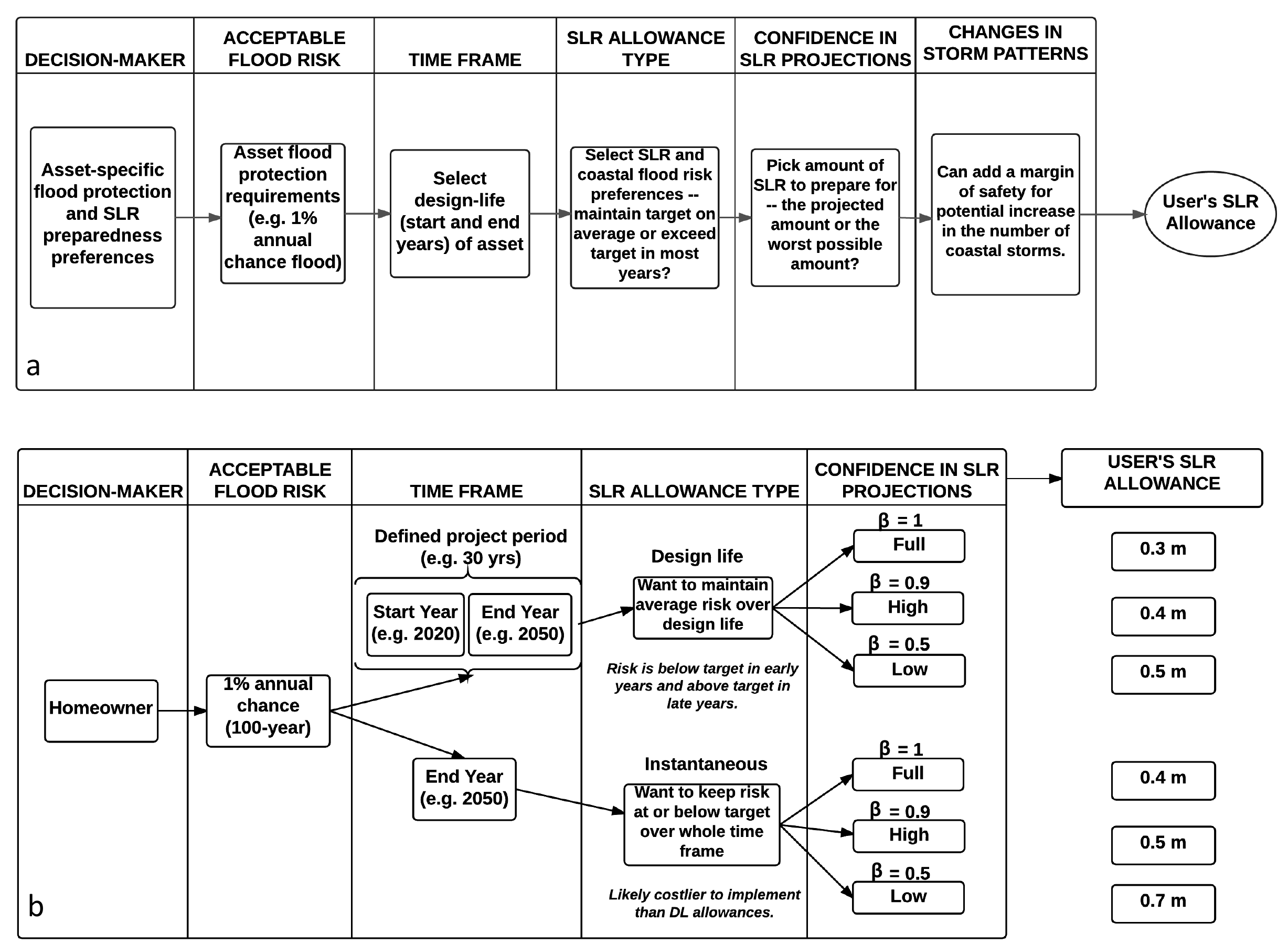} 
\caption{(a) A flow chart of the combined SLR allowance framework, and (b) a simple example of its application for a homeowner in Boston seeking to maintain 1\% AEP flood hazard over a mortgage from 2020 to 2050. See section 2.5.}
\end{center}
\end{figure}

\section{Results} 

%time when AEPs diverge (allowances figure). AdakIsland 2040-2060, Alameda 2050-2070, Anchorage 2030-2050, Annapolis 2050-2070. 2040 through 2070...instantaneous, show mean of 1% allowance, 10% allowance, and 0.02% allowance.

Across U.S. tide gauges, the instantaneous allowance $A$ is strongly correlated with expected sea-level rise $\E[\Delta_t]$ (Figure 3a). This is to be expected; as demonstrated in the SI, the offset between the instantaneous allowance $A(t)$ and the expected sea-level rise $\E[\Delta_t]$ does not depend on the first moment of the distribution of $\Delta_t$, although it does depend on higher-order moments and on the parameters of the extreme flood level distribution. (For example, for a zero-variance projection, the allowance is equal to the expected sea-level rise; increasing variance increases the allowance.) Accordingly, if the higher-order moments and extreme flood level distribution were identical across sites and only the expected sea-level rise differed, Figure 3a would show these points along a line with slope 1. Across all sites, the instantaneous allowance is larger than expected SLR on average by 4 cm in 2050 and 60 cm in 2100. This gap increases because the variance and skewness of the SLR projections increase over the course of the century. %Higher-order moments of the SLR distribution become increasingly important late in the century, reflecting the growing non-normality.

\begin{figure}[h!]
\includegraphics[width=7in]{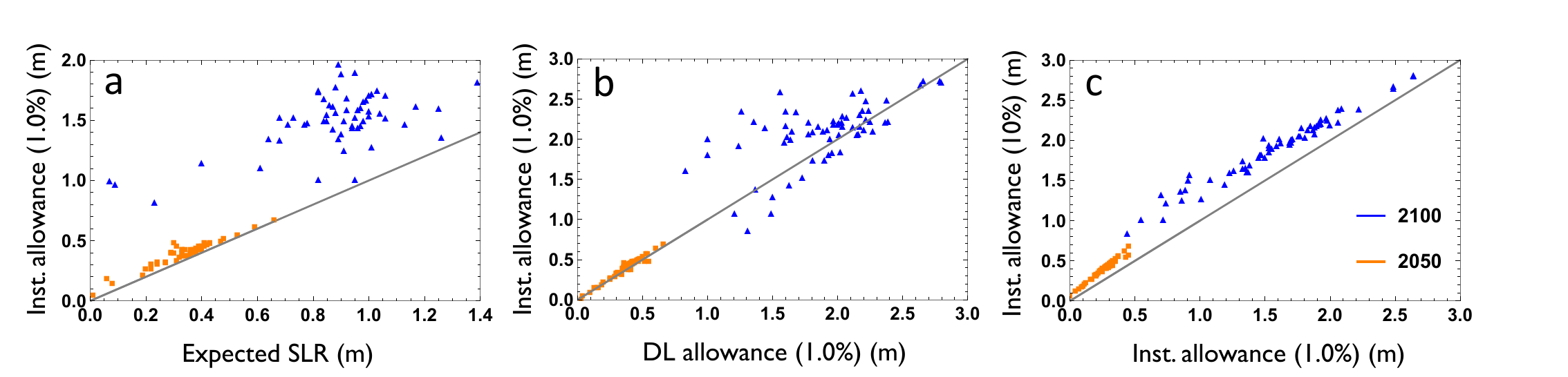} 
\caption{(a) 1\% instantaneous allowance vs. expected SLR (2050, 2100), (b) 1\% instantaneous allowance (2050, 2100) vs. 1\% DL allowance (2020-2050, 2020-2100) (c) 10\% instantaneous allowances vs. 1\% instantaneous allowance (2050, 2100). 2050 values are indicated by orange squares and 2100 values by blue triangles. All plots include a line with slope 1 for comparison. %(c) residuals of instantaneous SLR allowances versus expected SLR against y=x line (2050 in orange, 2100 in blue), and (d) residuals of integrated SLR allowances versus integrated SLR against y=x line (2050 in orange, 2100 in blue).
} 
\end{figure}

Figure 4 shows several flood return curves for Boston, Washington, D.C., and San Diego. First, it shows the historic flood curve ($N$), accounting for uncertainty in the GPD fit. Second, it shows flood curves adjusted for deterministic SLR estimates equal to the expected  value of SLR ($N+E(SL_t)$) and the worst-case SLR ($N+SL_{99.9}(t)$) in 2050 and 2100. Third, it shows the expected flood curves under the full PDF of SLR for 2050 and 2100 ($N_{e}(t))$). The instantaneous allowances for 2050 and 2100 are given by the horizontal offsets between the historic curve and the expected curves ($N_{e}(t))$). Fourth, the figure shows average expectations under the full PDF of SLR integrated over 2020--2050 and 2020--2100 design lives ($N_{e}(t_{1},t_{2})$); these curves are the AADLLs. The DL allowances are given by the horizontal offsets between the historic curve and the AADLL curves.

%,  at Boston. It also shows .. as an example (see SI for similar figures for all sites). While the flood return curve adjusted for projected SLR is an instantaneous flood level, the return curve associated with SLR integrated over a design-life is an AADLL. %Moderate flooding at Boston has been more common than extreme flooding, giving rise to a small, positive shape factor ($\xi = 0.074$). %and causing the flexure of the historic return level curve ($N$)---the log of the number of expected floods exceeding height $z$ per year. 
Accounting for uncertainty in SLR shifts the flood curves farther to the right than deterministically adjusting for the expected SLR, and therefore yields a curve significantly closer to the deterministic worst-case SLR scenario. The shifts of $N(z)$ by expected and projected SLR are not parallel; the range of uncertainty in SLR over time and $N$-dependence of the GPD alters the width of the shifts. The kinks in the figure arise at the transition in the extreme-value distribution between the extremes represented by the GPD and the extremes represented by a Gumbel distribution from $\lambda$ to 182.6 floods per year; a second kink arises at >182.6 floods per year (see Section 2.2). The appearance of these kinks in the $N_e(t)$ and $N_e(t_{1},t_{2})$ curves reflects the influence of high-end SLR projections that cause floods to transition between regimes.

\begin{figure}[h!]
\begin{center}
\includegraphics[width=5.7in]{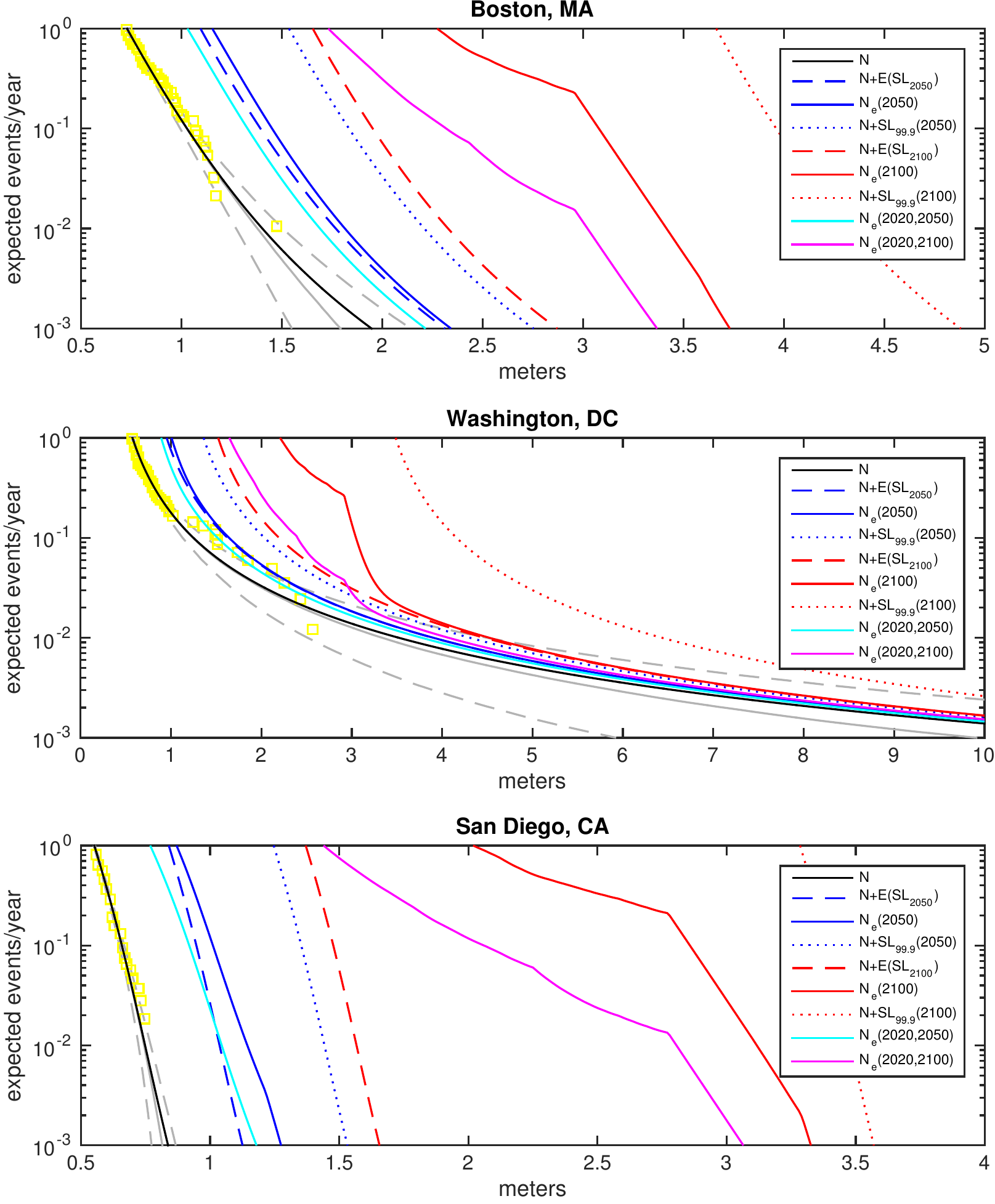} 
\caption{Flood return curves indicate the relationship between the number of expected flood events $N(z)$ and flood level $(z)$ for different assumptions of SLR, date, and time period. $N$ denotes the historic flood return curve, yellow points are empirical observations, grey lines are the 5th, 50th, and 95th percentiles of the GPD parameter uncertainty range. Fixed offsets of the historic curve for expected SLR in 2050 and 2100 are represented by $N + E(SL_{2050})$ and $N + E(SL_{2100})$, and 99.9th percentile SLR by $N + SL_{99.9}(2050)$ and $N + SL_{99.9}(2100)$. Instantaneous expected flood return levels for 2050 and 2100 are $N_{e}(2050)$ and $N_{e}(2100)$. AADLLs from 2020 to 2050 and from 2020 to 2100 are denoted as $N_{e}(2020,2050)$ and $N_{e}(2020,2100)$, with expected number of floods per year plotted as a function of flood height. The horizontal distances between the historic flood return levels and the $N_{e}(t))$/$N_{e}(t_{1},t_{2})$ represent the instantaneous/DL allowances, respectively.} 
%Historic flood return levels ($N$), fixed offsets of the historic curve for expected ($N+E(SL_{2050})$, $N+E(SL_{2100}$) and 99.9th percentile ($N+SL_{99.9}(2050)$, $N+SL_{99.9}(2100)$) SLR, instantaneous expected flood return levels ($N_{e}(2050)$, $N_{e}(2100)$), and AADLLs ($N_{e}(2020, 2050)$, $N_{e}(2020, 2100)$), with expected number of flood per year plotted as a function of flood height. The horizontal distances between the historic flood return levels and the instantaneous flood return levels/AADLLs  represent the instantaneous SLR/DL-SLR allowances.}
\end{center}
\end{figure}

%\begin{figure}[h!]
%\begin{subfigure}{.7\textwidth}
%  \centering
%  \includegraphics[width=.7\linewidth]{Boston-MA_returncurves}
%  \caption{}
%  \label{fig:sfig2}
%\end{subfigure} \\
%\begin{subfigure}{.7\textwidth}
%  \centering
%  \includegraphics[width=.7\linewidth]{Washington-DC_returncurves}
%  \caption{}
%  \label{fig:sfig1}
%\end{subfigure} \\
%\begin{subfigure}{.7\textwidth}
%  \centering
%  \includegraphics[width=.7\linewidth]{SanFrancisco-CA_returncurves}
%  \caption{}
%  \label{fig:sfig2}
%\end{subfigure} \\
%\caption{}
%\label{fig:fig}
%\end{figure}

Because the return levels of AEPs increase over time with SLR, the $p\%$ instantaneous allowance for year $t_2$ is always more conservative (higher) than the $p\%$ DL allowance over a period ending in year $t_2$ (Table 1, Figure 3b and SI). Adjusting historical flood levels upward by the instantaneous allowance at $t_2$ is akin to employing the MiniMax Flood Design Level \citep{Rootzen2013a}, which maintains a $p\%$ annual flood probability over every year in the project period (as opposed to averaged over the period). 

Among our representative set of 71 U.S. coastal tide gauges, allowances are nearly independent of $N$ in 2050, but significantly $N$-dependent by 2100 (Figure 3c). %add figure of cross plot of .002  and .1 allowance in 2050 and 2100 and maybe other time slices. and another one of the .1 and .01. they should have slope 1 if they are the same. have symbols for the different years...add to panel earlier.
Table 1 provides the 1\% instantaneous and DL allowances for every 21st century decade for various $\beta$ values (1, 0.9, 0.5, and 0) under RCP 8.5 for representative sites. Among all sites in the contiguous U.S. and Hawai`i, the 1\% instantaneous allowances for 2050 and 2100 have a mean and range of (0.37/ 0.03 to 0.66 m) and (1.89 / 0.83 to 2.80 m) with respect to the historic baseline, while the 1\% DL allowances for 2020-2050 and 2020-2100 have a mean and range of (0.27/ 0.04 to 0.45 m) and (1.53 / 0.14 to 2.64 m). These values demonstrate the gap between current flood risk protection standards and future flood return levels. Some Alaskan sites (Juneau, Seldovia, Seward, Skagway, Unalaska, and Yakutat) have negative SLR allowances arising from the projected falls in relative sea levels, due to a combination of glacial-isostatic adjustment, gravitational and flexural effects, and tectonics \citep{Kopp2014a}.

\begin{table}
\begin{center}
\caption{Vertical adjustments to infrastructure to maintain 1\% annual chance flood risk with projected SLR for a specific year (instantaneous) or over a design life. 1\% instantaneous and design-life allowances are in meters above the year 2000 baseline. DL allowances are integrated from 2020 to the specified year. $\beta$ values of 1, 0.9, 0.5, and 0 correspond to full, high, 50\%, and no confidence in the local SLR projection.} %title of the table
\begin{tabular}{c|c|c|c|c|c|c|c|c|c|c}
\hline
& \text{$\beta$} & 2020 & 2030 & 2040 & 2050 & 2060 & 2070 & 2080 & 2090 & 2100 \\
\hline
 \textbf{Boston} \\
 \hline
 \text{Instantaneous} & 1 & 0.13 & 0.21 & 0.3 & 0.41 & 0.55 & 0.76 & 1.07 & 1.50 & 2.01 \\
 \text{} & 0.9 & 0.16 & 0.25 & 0.36 & 0.53 & 0.79 & 1.15 & 1.58 & 2.05 & 2.56 \\
 \text{} & 0.5 & 0.22 & 0.34 & 0.49 & 0.70 & 1.01 & 1.40 & 1.83 & 2.30 & 2.81 \\
 \text{} & 0 & 0.28 & 0.41 & 0.58 & 0.81 & 1.13 & 1.52 & 1.95 & 2.42 & 2.93 \\
 \hline
\text{Design-life} & 1 & \text{} & 0.18 & 0.22 & 0.29 & 0.37 & 0.51 & 0.75 & 1.14 & 1.61 \\
 \text{} & 0.9 & \text{} & 0.21 & 0.27 & 0.37 & 0.56 & 0.87 & 1.27 & 1.71 & 2.21 \\
 \text{} & 0.5 & \text{} & 0.29 & 0.38 & 0.51 & 0.74 & 1.07 & 1.47 & 1.92 & 2.41 \\
 \text{} & 0 & \text{} & 0.36 & 0.46 & 0.60 & 0.84 & 1.17 & 1.57 & 2.01 & 2.5 \\
 \hline
 \textbf{Washington, D.C.} \\
 \hline
 \text{Instantaneous} & 1 & 0.13 & 0.21 & 0.29 & 0.38 & 0.49 & 0.61 & 0.73 & 0.86 & 1.00 \\
 \text{} & 0.9 & 0.14 & 0.23 & 0.32 & 0.43 & 0.58 & 0.74 & 0.93 & 1.17 & 1.47 \\
 \text{} & 0.5 & 0.19 & 0.3 & 0.43 & 0.60 & 0.85 & 1.14 & 1.47 & 1.86 & 2.31 \\
 \text{} & 0 & 0.25 & 0.39 & 0.56 & 0.79 & 1.13 & 1.50 & 1.92 & 2.38 & 2.91 \\
\hline
\text{Design-life}  & 1 & \text{} & 0.17 & 0.21 & 0.25 & 0.30 & 0.36 & 0.42 & 0.48 & 0.55 \\
 \text{} & 0.9 & \text{} & 0.19 & 0.23 & 0.28 & 0.34 & 0.42 & 0.50 & 0.61 & 0.76 \\
 \text{} & 0.5 & \text{} & 0.25 & 0.31 & 0.39 & 0.49 & 0.61 & 0.78 & 0.98 & 1.23 \\
 \text{} & 0 & \text{} & 0.32 & 0.40 & 0.50 & 0.64 & 0.82 & 1.04 & 1.29 & 1.60 \\
 \hline
 \textbf{Key West}  \\
\hline
\text{Instantaneous} & 1 & 0.11 & 0.18 & 0.26 & 0.38 & 0.61 & 0.94 & 1.36 & 1.83 & 2.38 \\
 \text{} & 0.9 & 0.13 & 0.21 & 0.32 & 0.53 & 0.83 & 1.18 & 1.59 & 2.06 & 2.60 \\
 \text{} & 0.5 & 0.17 & 0.27 & 0.42 & 0.67 & 0.98 & 1.33 & 1.74 & 2.22 & 2.76 \\
 \text{} & 0 & 0.21 & 0.32 & 0.48 & 0.75 & 1.06 & 1.42 & 1.84 & 2.31 & 2.85 \\
\hline
\text{Design-life} & 1 & \text{} & 0.15 & 0.20 & 0.28 & 0.48 & 0.81 & 1.21 & 1.68 & 2.22 \\
 \text{} & 0.9 & \text{} & 0.17 & 0.24 & 0.42 & 0.70 & 1.04 & 1.45 & 1.91 & 2.45 \\
 \text{} & 0.5 & \text{} & 0.23 & 0.32 & 0.52 & 0.80 & 1.14 & 1.53 & 1.99 & 2.53 \\
 \text{} & 0 & \text{} & 0.27 & 0.38 & 0.57 & 0.85 & 1.19 & 1.58 & 2.04 & 2.57 \\
 \hline
 \textbf{Grand Isle} \\
\hline
\text{Instantaneous} & 1 & 0.24 & 0.37 & 0.51 & 0.66 & 0.82 & 0.99 & 1.18 & 1.37 & 1.56 \\
 \text{} & 0.9 & 0.25 & 0.39 & 0.53 & 0.71 & 0.91 & 1.13 & 1.38 & 1.67 & 2.03 \\
 \text{} & 0.5 & 0.29 & 0.45 & 0.63 & 0.89 & 1.18 & 1.52 & 1.91 & 2.35 & 2.87 \\
 \text{} & 0 & 0.35 & 0.52 & 0.74 & 1.08 & 1.45 & 1.88 & 2.35 & 2.88 & 3.47 \\
 \hline
\text{Design-life} & 1 & \text{} & 0.31 & 0.38 & 0.45 & 0.53 & 0.62 & 0.71 & 0.81 & 0.92 \\
 \text{} & 0.9 & \text{} & 0.32 & 0.39 & 0.48 & 0.57 & 0.68 & 0.8 & 0.96 & 1.18 \\
 \text{} & 0.5 & \text{} & 0.37 & 0.46 & 0.57 & 0.71 & 0.88 & 1.09 & 1.34 & 1.67 \\
 \text{} & 0 & \text{} & 0.44 & 0.54 & 0.68 & 0.86 & 1.09 & 1.35 & 1.66 & 2.04 \\
 \hline
 \textbf{San Diego}  \\
\hline
 \text{Instantaneous}  & 1 & 0.09 & 0.15 & 0.24 & 0.39 & 0.64 & 0.98 & 1.39 & 1.85 & 2.37 \\
 \text{} & 0.9 & 0.11 & 0.18 & 0.35 & 0.60 & 0.91 & 1.27 & 1.68 & 2.13 & 2.64 \\
 \text{} & 0.5 & 0.14 & 0.23 & 0.41 & 0.67 & 0.98 & 1.33 & 1.74 & 2.19 & 2.70 \\
 \text{} & 0 & 0.16 & 0.25 & 0.44 & 0.69 & 1.00 & 1.35 & 1.76 & 2.22 & 2.73 \\
 \hline
\text{Design-life}  & 1 & \text{} & 0.13 & 0.19 & 0.30 & 0.46 & 0.74 & 1.12 & 1.56 & 2.06 \\
 \text{} & 0.9 & \text{} & 0.16 & 0.29 & 0.52 & 0.82 & 1.16 & 1.56 & 2.01 & 2.51 \\
 \text{} & 0.5 & \text{} & 0.20 & 0.36 & 0.60 & 0.89 & 1.24 & 1.64 & 2.08 & 2.59 \\
 \text{} & 0 & \text{} & 0.23 & 0.38 & 0.62 & 0.92 & 1.27 & 1.67 & 2.11 & 2.62 \\
\hline
\end{tabular}
\end{center}
\end{table}

\begin{figure}[h!]
\begin{center}
\includegraphics[width=6.5in]{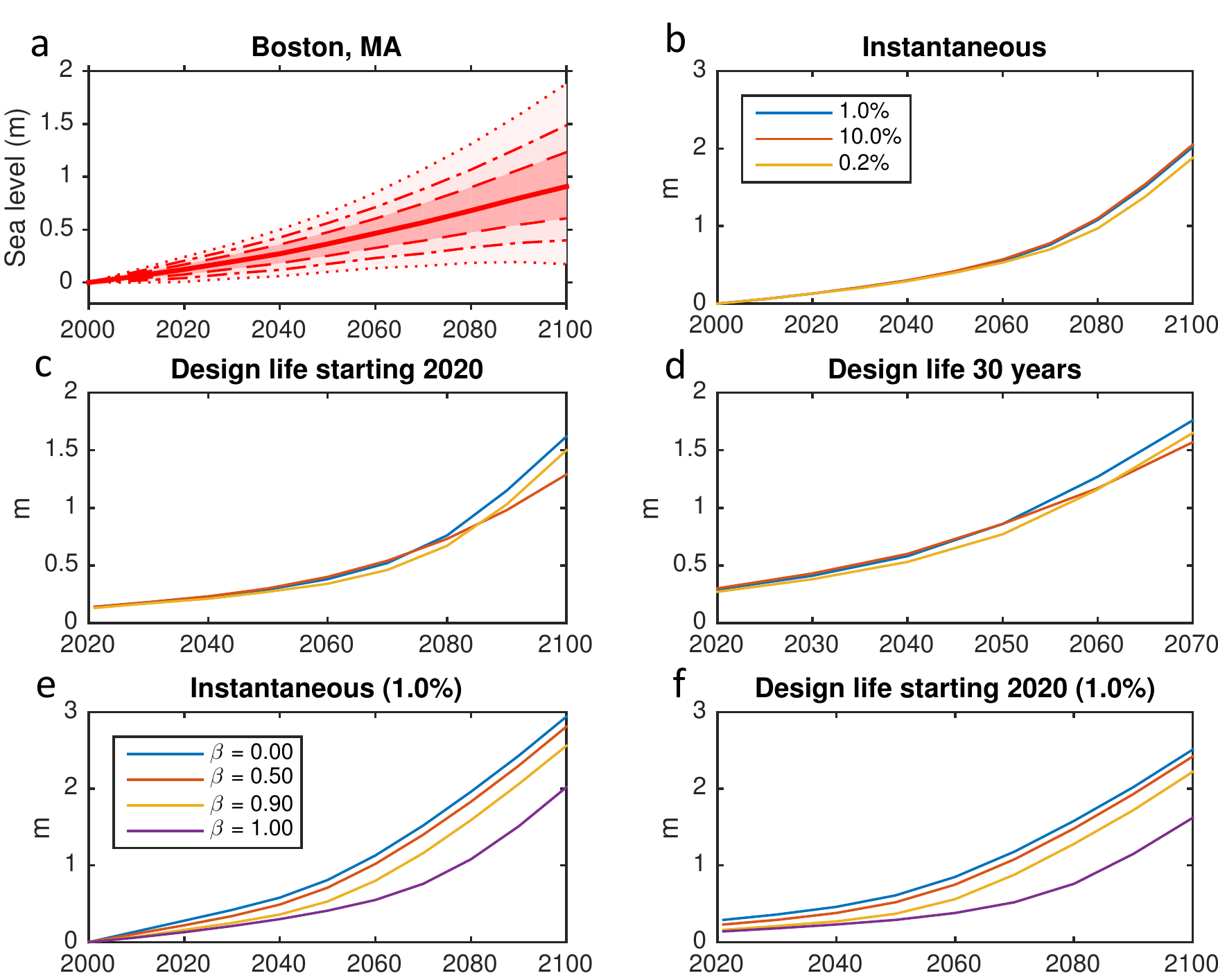} 
\caption{(a) Local SLR projections for Boston, (b) instantaneous allowances for various risk levels ($N_0$ = 10\%,1\%, and 0.2\% AEP), (c) DL allowances starting in 2020 for variable project lengths (from 1 to 81 years), (d) DL allowances for 30-year projects with variable start dates (from 2020 to 2070), (e) 1\% instantaneous allowances with Limited Degree of Confidence (LDC) metric and (f) 1\% DL allowances with LDC metric.}
\end{center}
\end{figure}

Figure 5 illustrates different instantaneous and DL allowances for an asset's risk tolerance (1\%,10\%, and 0.2\% annual chance of flooding) and time period, local sea-level rise projection and confidence therein. For example, in Boston, there is little difference between the 10\%, 1\%, and 0.2\% allowances before late century (Figure 5b-d). However, allowance amounts are sensitive to the time period of a project; a 30-year 1\% risk tolerant asset will have an allowance twice as large if starting in 2050 rather than 2030 (Figure 5d). As noted, SLR allowances will be lower for decision-makers with full confidence in the expert SLR PDF than those with full confidence in worst-case SLR. Because of the approximately log-linear relationship between $N$ and $z$, the worst-case possibility exerts a strong influence on allowances even for high degrees of confidence. For example, for an 80-year asset, the 1\% DL allowance for 2020--2100 is 1.6 m with full confidence in the expert PDF ($\beta = 1.0$) and 2.5 m with no confidence in the expert PDF ($\beta = 0$) (Figure 5f). Due to the pull of the high end of the local SLR projection, $\beta = $0.92 yields a DL allowance that is halfway between the metrics of full confidence and a complete lack of confidence in projected SLR (i.e. no-confidence allowance, $\beta = 0$) and $\beta = 0.5$ yields a DL allowance of 2.4 m, quite close to the no-confidence allowance. Tables and figures analogous to Table 1 and Figures 4-5 of flood levels and allowances for all sites are available in the SI. 

%For example, for Boston, the DL allowance over 2020--2050 (0.29 m) is less than the instantaneous allowance in 2050 (0.41 m), and the DL allowance over 2020--2100 (1.61 m) is less than the instantaneous allowance in 2100 (2.01 m). 

%As uncertainty in the local SLR PDF grows with time as SLR commitments unfold throughout the century, long-term allowances---and particularly instantaneous allowances, which are driven by late-century SLR---cannot be predicted from expected or integrated SLR alone. 

\section{Discussion}
%Our allowance framework is part of a larger effort to provide robust climate science to the decision-making level. Our findings emphasize that the notion of the historic 100-year flood as a convention is truly outdated and should be replaced by a more robust local indicator of dynamic flood risk, such as the entire probability distribution of flood return periods. As DL-SLR allowances take temporal dynamics of SLR into account, they can be used to account for non-stationary flood risk while retaining current flood risk tolerance levels in emerging flood protection standards.

%It is important to note that the historic 1\% AEP flood height metric (e.g., 100-year flood)---the predominant flood risk metric used by the National Flood Insurance Program---is essentially a convention which assumes near-perfect protection against smaller, more frequent flood events. Assuming a stationary distribution of flood events with shorter return periods provides the perception of protection against the combined risk from the accumulation of the 100-year and more-frequent floods (such as the 50-year flood). The convention also assumes a stationary distribution of longer return period flood events (such as the 500-year flood)---and some protection against them. As shown by the AADLL and instantaneous flood levels (Figure 4), that assumption no longer holds because the expected distribution of all flood return periods changes with SLR. Thus, the scaling of return periods for both previously more and less frequent events is now of great importance for decision-making. 

It is important to note that the historic 1\% AEP flood height (e.g., 100-year flood)---the predominant flood risk metric used by the National Flood Insurance Program---is essentially a convention which assumes near-perfect protection against flood events. As shown by the AADLL and instantaneous flood levels (Figure 4), that assumption no longer holds because the expected distribution of all flood return periods changes with SLR. Complementing other methods, the framework presented here is part of a larger effort to provide robust climate science to the decision-making level \citep[e.g.,][]{Jonkman2009a, Lempert2012a} (see SI). SLR allowances provide a means for stakeholders to account for the distribution of flood return levels and maintain a desired protection standard---such as 1\% AEP---under non-stationarity. %Our allowance framework is part of a larger effort to provide robust climate science to the decision-making level. Complementing other methods, the framework presented here has a combination of traits to allow for some ease in transition from stationary to non-stationary flood risk management (see SI).  

The framework has a combination of traits to allow for some ease in transition from stationary to non-stationary flood risk management. First, its sea-level rise estimates include local factors. While global mean sea level (GMSL) change is mainly driven by land-ice melt and by thermal expansion of warming ocean water, ocean/atmosphere dynamics, static-equilibrium sea-level fingerprints, and other regional factors contribute significantly to local sea-level change \citep{Kopp2015a}. Second, its sea-level rise estimates are based on complete probability distributions, as opposed to central estimates, scenarios with unspecified probabilities, or likely ranges. Third, its allowances reflect decision-makers' risk management preferences, such as desired protection level, limited degree of confidence in SLR projections, and preferences between protection and cost which relate to the choice of instantaneous vs. DL allowances. Accounting for different flood risk tolerance levels is critical, as households, businesses, and government entities often have divergent risk perceptions and behavior \citep{Willis2007a}, although this generally seems to exert a minor influence on allowances before mid-century. Fourth, its allowances account for different planning periods throughout the 21st century, capturing the effects of SLR over time. Finally, similar to the majority of previous methods, the framework relies on the historic distribution of storms, as future projections are not yet well understood and only available for a few tide gauges \citep[e.g.,][]{Lin2012a}.

A ``project'' can be any investment time period, such as a 30-year mortgage or 80-year power generation facility. Given a planner's acceptable flood risk (e.g., 1\%, 10\%, 0.2\% AEP), she can identify the corresponding protection height during a planning horizon, such as a 30-year project from 2030-2060 or an 80-year project from 2020-2100 for various levels of confidence in SLR projections (Figure 5). Decision-makers can explore and adjust variables they have control over, such as potential implementation delays, \textit{a priori}. Stochastic planning of this type is particularly important for large capital projects, especially those with lengthy lifetimes, to achieve practical maintenance expectations and avoid risk tolerance exceedances. For example, allowances can be used to explore the effect of potential delay of bridge construction or post-mortgage occupation of a house on protection height or risk taken. Similarly, the framework can inform rational thinking about trade-offs between flexibility (in terms of adding protection over time) and regrets (in terms of overprotection) in adaptation strategies. In this regard, AADLL and instantaneous flood levels can be integrated with flexible adaptation pathways \citep{Haasnoot2013a} and to help identify dates when acceptable flood risk is crossed \citep{Kwadijk2010a}. Instantaneous flood levels can be used as an upfront high fixed cost adaptation strategy, which may be inflexible over time \citep{Ranger2013a}. AADLLs can be used to inform terminal adaption strategies that at a certain date should either be upgraded (by adding more freeboard associated with a revisited design life) or in transition to another adaptation strategy (and abandoning the asset).

There is an implicit trade-off between instantaneous and DL allowance types in terms of flood protection and cost. Project end-year instantaneous allowances are below the target annual risk level (and therefore lead to excess protection) in all but the last design-life year, but by requiring a larger freeboard, may be costlier to implement than DL allowances. For example, to maintain 1\% flood risk tolerance for an asset from 2020 to 2080, the DL and instantaneous allowances are 0.7 m and 0.4 m in Washington, D.C., respectively (Table 1). Raising infrastructure (or its flood defense) by the additional 0.3 m could cost \$5,000 (USD) \citep{USACE2015a}. %---whereby protection cost estimates are time and location dependent. 
The increased protection cost may be preferred for projects where the lifetime of interest may extend well beyond the nominal design life or where uncertainty about the extent of SLR is very high.  %As DL allowances keep annual risk below target in early years and above target in late years, 
Conversely, DL allowances may be preferred for a well-defined design life (e.g., such as a 30-year mortgage) where minimal value is imputed to flooding after the end of the design life. As DL allowances provide a minimum freeboard to maintain desired protection on average over the project's design life, they may also be preferred as a low-regrets options when financial resources for resilience are scarce, which is a common barrier for adaptation \citep[e.g.,][]{Moser2010a}. %(Although investing in conservative protection may seem more logical to some, it is not often practiced in reality. For example, the issue of upfront cost led to reducing protection below the federal 1\% AEP protection standard in New Orleans pre-Katrina \citep{Kunreuther2006a, Burby2006a}). 

In some states in the U.S., political leaders have been reluctant to discuss the human-caused acceleration of SLR \citep[e.g.,][]{Kopp2015b}. However, while a relatively modest lack of confidence in expert PDFs toward the worst-case possibility can dominate the calculation of allowances, even a modest degree of confidence allows the expert PDFs to dominate. To illustrate, we calculate a variant of the LDC allowances wherein limited confidence in the expert PDFs is expressed by belief not in the worst case scenario, but in zero sea-level rise. To distinguish these Panglossian Limited Degree of Confidence from the more traditional LDC metric, we use values of $\beta'$, i.e., 
\begin{equation}
N_{e,PLDC}(z,t,\beta') = \beta' N_{e}(\Delta, t) + (1-\beta') N(z) 
\end{equation}
A party must be very optimistic and have quite low confidence in expert SLR PDFs to argue against preparing for SLR.
For example, $\beta'= 0.5$ means a party believes there is a 50\% chance that the expert PDF is correct a 50\% chance that there will be no SLR. Even if they think there is a 90\% chance of no sea-level rise and a 10\% chance the experts are correct ($\beta' = 0.1$), the appropriate allowance over 2020-2100 in Boston for example is still about half the allowance as if they had full confidence in the expert PDF ($\beta'=1$), rationalizing adaptation planning. %The difference between these allowances lessens over time.

Risk can be defined as the probability of an event's occurrence and the consequence of its impact \citep{Lavell2012a}. From the perspective of a coastal decision-maker, SLR allowances capture the changing probability of an event's occurrence (and the number of annual exceedances). Although they do not directly provide information regarding the consequence of impact, by holding the decision-maker's risk tolerance constant, allowances provide an adaptation offset to counteract the adverse consequence of exceeding tolerable risk. The framework can be coupled with damage functions that account for the consequence of flooding. Accompanied by such, the allowances can provide a mechanism to translate comprehensive SLR PDFs into actionable science to meet local adaptation risk management decisions sensitive to extreme events. %Our analysis assumes a binary step fragility function, with no concern about flooding below the flood level of interest. 

We focus on flood height which is a primary metric decision-makers what to know \citep{Jonkman2009a, Neumann2010a, Lempert2012a, Woodward2013a}. Other hydraulic factors, such as surge duration, are important in assessing inundation and can be incorporated in future work. SLR allowances can also be considered holistically with other coincident hazards in our changing climate (such as riverine flooding or extreme precipitation), which are developing research areas \citep{Wahl2015a, Katsman2011a}. Moreover, learning to better accommodate with overtopping of defenses is critical in a non-stationary climate \citep{Brown2010a}. Beyond methodology, accounting for SLR and its uncertainty in federal and municipal flood standards also requires institutional changes, which have proven to be an obstacle for effective risk management \citep[e.g.,][]{Moser2010a}.

Finally, our model is a `bath tub' model in that it accounts for mean wave height, which is often but not always a good approximation \citep{Lin2012a, Georgas2014a}. We have assumed a historic distribution of storms, which imperfectly samples the true probability distribution which may change in a warming climate \citep{Christensen2013a}. Projection of changes in storminess involves deeper layers of uncertainty and is a nascent area of research for individual basins \citep{Christensen2013a}.  Users with a high risk aversion to potential changes in storminess may also include an additional margin of safety. %For improved accuracy, hydrodynamical modeling \citep[e.g.][]{Lin2010b, Lin2012a} may be required in areas with high exposure to tropical cyclones \citep{Tebaldi2012a}. 

%\begin{figure}[h!]
%\begin{center}
%\includegraphics[width=6.5in]{Boston-MA_Allowances_negativebeta.pdf} 
%\caption{Instantaneous (left) and DL (right) SLR allowances calculated with Panglossian Limited Degrees of Confidence. For Panglossian LDCs, $\beta$- denotes the degree of confidence in the expert PDF and $1+\beta$ the degree of belief in no SLR.}
%\end{center}
%\end{figure}

\section{Conclusions}
The availability of probabilistic local SLR projections provides an opportunity to improve coastal flood risk management. In this study, we provide a framework for local, dynamic, and actionable flood hazard information that can be used by stakeholders to inform flood risk management despite ambiguity in SLR projections. Our calculations of average annual design-life flood levels, instantaneous allowances and design-life allowances illustrate the importance of accounting for asset specific time frames and deep uncertainty in SLR projections to satisfy project design standards and risk preferences. Because of the evolution of flood levels in a non-stationary climate, failing to do so can compromise standards of protection, even from short project delays or extended durations. In this effort to provide individuals with actionable climate science, households, businesses, and government entities can select a SLR allowance that meets their planning needs among trade-offs, such as between protection and adaptation cost, and between flexibility and regret. %Moreover, tolerable number of events above a given level during a period and confidence in sea-level rise projections are important but not exhaustive measures of the information needed for local decisions. 
The potential severity of exceedance resulting from deeply uncertain changing storm dynamics, such as hurricane intensity, also matters and needs to be better accounted for in future local flood risk management. To summarize, our work underscores the need to readjust federal and local planning beyond the historic 100-year flood to an adaptable means of maintaining flood risk standards, such as that afforded by design-life and improved instantaneous allowances.

\section{Acknowledgements}
This publication is the result of research sponsored by the New Jersey Sea Grant Consortium (NJSGC) (publication NJSG-16-898) with funds from the NJSGC and National Oceanic and Atmospheric Administration grant NA14OAR4170085. C. Tebaldi was supported by the Regional and Global Climate Modeling Program (RGCM) of the U.S. Department of Energy's, Office of Science (BER), Cooperative Agreement DE-FC02-97ER62402. The statements, findings, conclusions, and recommendations are those of the authors and do not necessarily reflect the views of the funding agencies.

%\bibliographystyle{spbasic_etal}
%\bibliography{DLSLRfoo}

\end{document}